\documentclass{iopjournal}

\usepackage{amsmath}
\usepackage{amssymb}
\usepackage{enumitem}
\usepackage{color}
\usepackage{braket}
\usepackage[normalem]{ulem}

\begin{document}

\articletype{Paper} 

\title{Isotropic random walks and Brownian diffusion on complex projective space}

\author{Gyula I. T\'oth$^1$\orcid{0000-0002-8446-9572}}

\affil{$^1$Department of Mathematical Sciences, Loughborough University, Loughborough, LE11 3TU, United Kingdom}

\email{g.i.toth@lboro.ac.uk}


\begin{abstract}
We study isotropic random walks on complex projective space in the context of stochastic state-space exploration in open quantum systems. Using results from harmonic analysis on compact rank-one symmetric spaces, first we obtain explicit transition kernels and characterise the evolution of fidelity and geometric observables. Then, we establish the correspondence between the small-step limit and Fubini–Study Brownian motion, and also show that the long-time limit reproduces Haar-random state statistics. At the level of the density matrix, the ensemble evolution formally  coincides with an isotropic depolarizing channel at all times. Our main results concern first-passage statistics to high-fidelity target states. In the Brownian limit, we derive a closed-form expression for the mean first-passage time and an explicit spectral representation of the first-passage-time distribution, with decay rates determined by the eigenvalues of the killed Fubini-Study Laplace-Beltrami operator. The resulting asymptotics reveal a strong dependence on target fidelity and Hilbert-space dimension, demonstrating the rapidly increasing difficulty of reaching high-fidelity targets through unbiased isotropic exploration. The model is purely geometric and does not incorporate microscopic Hamiltonians or Lindblad dynamics. Nevertheless, the results suggest that it offers a natural starting point for stochastic models of open quantum systems that incorporate Hamiltonian and dissipative effects, with potential applications to stochastic quantum state preparation and control.
\end{abstract}

\section{Motivation}

Designing next-generation quantum devices remains a major challenge due to the difficulty of modelling quantum decoherence, a physical process in which interactions with the environment suppress quantum interference and thereby limit device performance \cite{BreuerPetruccione2002,Zurek2003}. As the combined device--environment system evolves in an exponentially large Hilbert space, direct microscopic Schr\"odinger simulations are computationally intractable on practical scales. Therefore, open quantum systems are commonly described in terms of reduced system dynamics obtained by tracing over environmental degrees of freedom. Under suitable weak-coupling and Markovian approximations, this leads to Lindblad master equations governing the deterministic evolution of reduced density matrices \cite{Lindblad1976,Gorini1976}. In practice, quantum trajectory theory gives a re-formulation of Lindblad dynamics in which the reduced density matrix is represented as the ensemble average over stochastic pure-state trajectories called ``unravelings''. In the quantum-jump (or Monte Carlo wave-function) formulation, the state evolves deterministically between stochastic jumps \cite{Dalibard1992,Dum1992}, while diffusive unravelings (such as continuous-measurement formulations \cite{JacobsSteck2006,Wiseman2009} or Quantum State Diffusion theory \cite{GisinPercival1992,Percival1998QSD}) lead to continuous-time stochastic Schr\"odinger equations describing diffusive pure-state evolution. Both jump-type and diffusive unravelings define Markov processes on the device's Hilbert space \cite{BreuerPetruccione2002,BarchielliGregoratti2009}. These approaches have proved highly successful both conceptually and computationally, and they have become standard tools for modelling dissipative many-body systems \cite{Daley2014}. However, as the transition structure of quantum trajectories depends explicitly on the Hamiltonian, Lindblad operators, and measurement scheme, the analytical tractability of the resulting -- generally non-linear and anisotropic -- stochastic evolution is often poor (except in highly symmetric or low-dimensional cases).

\section{The model}

The limitations of quantum trajectory theory discussed above motivate the present work. Following Watabe et al. \cite{Watabe2022}, we propose an analytically tractable baseline model for stochastic pure-state evolution based on isotropic random walks on complex projective space. By construction, the model does not explicitly incorporate Hamiltonians or Lindblad operators and is therefore not intended to provide a microscopic description of a specific open quantum system. Instead, it offers an effective geometric framework for studying stochastic exploration of quantum-state space, while extensions are expected to be necessary to describe specific physical systems.  

The geometric setting of the proposed model is as follows. In standard practice, the state vector $\ket{\psi}$ representing a pure state of the reduced system (i.e., the system without the environmental degrees of freedom) is an element of the complex Hilbert space $\mathcal{H}$. Due to the normalisation condition, pure states may be represented as points on the unit sphere $S \equiv \{ u \in \mathcal{H} : \lVert u \rVert =1 \}$, where $\|.\|=\sqrt{\langle .,. \rangle}$ is the standard norm and $\langle ,.,\rangle$ is the conjugate-linear inner product (the Dirac notation will be omitted in most of the paper henceforth). Two vectors $u,v\in S$ represent the same physical state whenever they differ only by a global phase, i.e.\ when $v=e^{\mathrm{i}\varphi}\,u$ for some $\varphi\in\mathbb{R}$; equivalently, $x,y\in\mathcal{H}\setminus\{\mathbf0\}$ represent the same physical state whenever $y=c\,x$ for some $c\in\mathbb{C}\setminus\{0\}$. This equivalence relation defines the projective Hilbert space (or ray space) $\mathbf{P}(\mathcal{H})$, whose elements correspond uniquely to pure quantum states \cite{BrodyHughston2001}. The natural metric on $\mathbf{P}(\mathcal{H})$ is given by the Fubini--Study distance \cite{BengtssonZyczkowski2006}:
\begin{equation*}
d(\xi,\eta)\equiv d_{FS}(x,y)=\cos^{-1}\left|\left\langle\frac{x}{\lVert x\rVert},\frac{y}{\lVert y\rVert}\right\rangle\right|\in[0,\pi/2]\,,
\end{equation*}
where $x,y\in \mathcal{H}$ are representatives of $\xi,\eta \in \mathbf{P}(\mathcal{H})$, respectively. Since $d_{FS}(x,y)>0$ only when $x$ and $y$ lie on distinct complex lines, the Fubini--Study metric provides the natural Riemannian geometry of the projective Hilbert space \cite{BengtssonZyczkowski2006}. For $\dim_\mathbb{C}(\mathcal{H})=n < \infty$, $\mathcal{H}$ is isomorphic to $\mathbb{C}^n$, and therefore $\mathbf{P}(\mathcal{H})$ is isomorphic to the complex projective space $\mathbb{CP}^{n-1}$. Geometrically, $\mathbb{CP}^{n-1}$ can be represented either as a Hopf fibration of the unit sphere in $\mathbb{C}^n$ or as a compact, rank one symmetric space:
\begin{equation*}
\mathbb{CP}^{n-1}\cong \frac{S^{2n-1}}{S^1} \cong \frac{SU(n)}{S(U(n-1)\times U(1))} \,.
\end{equation*}
In the Monte Carlo wave-function formulation, the state undergoes deterministic non-unitary evolution interrupted by ``quantum jumps'' at stochastic times. At each jump event, the system selects a pure state according to classical probability laws. This structure motivates considering simplified stochastic models defined directly on the projective space. The general framework is a discrete stochastic process $\xi_m$ of pure states on $\mathbb{CP}^{n-1}$, where $\xi_0$ is the fixed initial state and the transition probabilities encode the stochastic jump dynamics. The present framework adopts the stochastic-geometric viewpoint at the level of an effective Markov process on projective Hilbert space and considers the maximally symmetric process compatible with the geometry of $\mathbb{CP}^{n-1}$. In this approach, temporal intervals between quantum jumps are replaced by distances on $\mathbb{CP}^{n-1}$, and the transition probability density from $\xi_{m-1}$ to $\xi_m$ is determined by a radial kernel.

With this geometric and dynamical setting in place, we emphasize that the mathematical ingredients of the construction are classical. The novelty of the present work lies in combining them within a single analytically tractable framework on projective Hilbert space, in which the isotropic random walk, its Fubini--Study Brownian limit, and the associated first-passage problem can be treated within the same construction. Accordingly, the structure of the rest of the paper is as follows. In Section 3, we review the general theory of isotropic random walks on compact rank one symmetric spaces. We derive explicit expressions for the radial transition probabilities and for geometric observables associated with the generating metric. In Section 4, we specialise the formalism to $\mathbb{CP}^{n-1}$. We first analyse the fidelity statistics and ensemble relaxation, which is followed by the analysis of the short-time diffusion and long-time geometric equilibration regimes. Here we establish the connection with Brownian motion generated by the Fubini--Study Laplace--Beltrami operator and derive asymptotic expressions for the geometric spreading of the walk (Section~4.2). In Section 4.3, we present the main analytical results of the paper: we formulate the first-passage problem in the Brownian limit, derive a closed-form expression for the mean first passage time and an explicit spectral representation of the first-passage-time distribution, and obtain the high-fidelity asymptotics of the principal decay rate. In Section 5, we validate our results by comparing them to classical results from quantum information theory and recent experimental results for specific open quantum systems. Section~6 contains concluding remarks, including the discussion of potential extensions of the model for real physical systems.

\section{Isotropic random walks on compact rank one symmetric spaces}

Let $V$ be a compact, rank one symmetric space with corresponding Jacobi parameters $\alpha$ and $\beta$. (This family consists of the unit sphere and $4$ projective spaces: the real projective space, the complex projective space, the quaternionic projective space, and the Cayley plane.) The algebra of radial functions on $V$ carries the Jacobi hypergroup, which emerges from harmonic analysis \cite{Koornwinder1973,Helgason1984,Pap_Voit_1998,Roesler_Voit_1999}. A direct consequence of this result is that the transition probability density of the isotropic random walk can be written in closed form on these spaces. We present the emerging formulae in their natural generality, since the $(\alpha,\beta)$ structure reveals symmetries that would be obscured by premature specialisation. The probability that the walker reaches the infinitesimally small neighbourhood of state $\xi \in \mathbb{CP}^{n-1}$ from initial state $\xi_0 \in \mathbb{CP}^{n-1}$ in exactly $m \in \mathbb{N}$ steps is given by $f_m(\xi|\xi_0)\,d\nu(\xi)$, where the Haar measure $d\nu(\xi)$ is normalised as $\int_V d\nu(\xi) =1$, while the $m$-step transition probability density $f_m(\xi|\xi_0)$ is a radial function, i.e., it only depends on the distance of the initial and final states. Accordingly, 
\begin{equation*}
f_m(\xi|\xi_0)= g_m(\vartheta) \,,
\end{equation*}
where $\vartheta=d(\xi_0,\xi)$, and $d(\xi,\eta)= d_{V}(x,y)$ $\in$ $[0, \pi/2]$ is the generating metric of $V$ ($x$ and $y$ are representatives of $\xi$ and $\eta$, respectively, in the embedding space). Harmonic analysis provides
\begin{equation}
\label{trans}
g_m(\vartheta) = \sum_{k=0}^\infty \pi_{\alpha,\beta}(k)\,\hat{\mu}_k^m J_k^{\alpha,\beta}(\cos(2\vartheta)) \,, 
\end{equation}
where $\pi_{\alpha,\beta}(k) =(2\,k+\alpha+\beta+1)\,(\alpha+\beta+1)_k(\alpha+1)_k/[(\alpha+\beta+1)\,k!\,(\beta+1)_k]$ stands for the Plancherel measure, and $J_k^{\alpha,\beta}(z)={}_2F_1(-k,k+\alpha+\beta+1,\alpha+1,(1-z)/2)$ is the normalised Jacobi polynomial. The single-step transition kernel for harmonic mode $k$ is the Jacobi-Fourier transform of the step-size probability distribution $d\mu(\vartheta)$:
\begin{equation*}
\hat{\mu}_k = \int_0^{\pi/2} J_k^{\alpha,\beta}(\cos(2\vartheta))\,d\mu(\vartheta) \,.
\end{equation*}
\paragraph{Proposition \cite{Pap_Voit_1998}.}
Let the step-size distribution be non-degenerate at $\vartheta=0$ and $\vartheta=\pi/2$. Then the isotropic random walk on a compact rank one symmetric space converges weakly to the uniform measure as $m\to\infty$. In fact, for any such $d\mu(\vartheta)$, $\hat{\mu}_0 = 1$ and $|\hat{\mu}_k|<1$ for $k \in \mathbb{N}$, and therefore only the $k=0$ mode survives for $m \to \infty$ in Eq. (\ref{trans}). This implies 
\begin{equation}
\label{Haareq}
\lim_{m \to \infty}f_m(\xi|\xi_0) = 1 \,,
\end{equation}
i.e., for any initial state, the final state becomes uniformly distributed on $V$ with respect to the normalised Haar measure as the number of steps diverges.

To obtain the radial transition probability density, we start from the normalised radial Haar measure on $V$, which is a Lebesgue measure:
\begin{equation*}
d\omega(\vartheta) = 2\,C_{\alpha,\beta}\,\sin^{2\alpha+1}(\vartheta) \, \cos^{2\beta+1}(\vartheta)\,d\vartheta \,,
\end{equation*}
where $C_{\alpha,\beta}= \Gamma(\alpha+\beta+2)/[\Gamma(\alpha+1)\,\Gamma(\beta+1)]$ is a normalisation factor, yielding $\int_0^{\pi/2}d\omega(\vartheta) =1$. The transition probability density is normalised as $\int_V f_m(\xi|\xi_0)\,d\nu(\xi)=1$ for every initial state $\xi_0$ and number of steps $m \in \mathbb{Z}_{\geq 0}$. Using $1$ $=$ $\int_V f_m(\xi|\xi_0)\,d\nu(\xi)$ $=$ $\int_0^{\pi/2} g_m(\vartheta)\,d\omega(\vartheta)$ $=$ $\int_0^{\pi/2} \mathcal{F}_m(\vartheta)\,d\vartheta$ then yields the radial $m$-step transition probability density
\begin{equation}
\label{pdf}
\mathcal{F}_m(\vartheta) = 2\,C_{\alpha,\beta}\,\sin^{2\alpha+1}(\vartheta)\cos^{2\beta+1}(\vartheta) \sum_{k=0}^\infty \pi_{\alpha,\beta}(k)\,\hat{\mu}_k^m\,J_{k}^{\alpha,\beta}(\cos(2\vartheta)) \,.
\end{equation}
The expectation of $\vartheta$ reads as:
\begin{equation*}
\mathbb{E}_m(\vartheta) = \int_0^{\pi/2} \vartheta\,\mathcal{F}_m(\vartheta)\,d\vartheta = \sum_{k=0}^\infty \pi_{\alpha,\beta}(k) \hat{\mu}_k^m\,E_k^{\alpha,\beta} \,,
\end{equation*}
where 
\begin{equation*}
E_k^{\alpha,\beta} = \int_0^{\pi/2} \vartheta\,J_k^{\alpha,\beta}(\cos(2\vartheta))\,d\omega(\vartheta) = C_{\alpha,\beta} \sum_{j=0}^k \frac{(-k)_j (k+\alpha+\beta+1)_j}{(\alpha+1)_j\,j!}\,I_{\alpha+j,\beta} \,,
\end{equation*}
and
\begin{equation*}
I_{r,\beta} = B\!\left(r + \frac{3}{2}, \beta + 1\right)
\,{}_3F_2\left(\left\{\frac{1}{2}, \frac{1}{2}, r + \frac{3}{2} \right\},\left\{\frac{3}{2}, r + \beta + \frac{5}{2}\right\},1\right) \,,
\end{equation*}
where $B(z_1,z_2)$ is the beta function, and ${}_3F_2(\{a_1,a_2,a_3\},\{b_1,b_2\},z)$ is the generalised hypergeometric function. When $m \to \infty$, only the $k=0$ mode contributes, thus yielding:
\begin{equation}
\label{limtheta}
\lim_{m \to \infty} \mathbb{E}_m(\vartheta) = E_0^{\alpha,\beta} = \frac{
B\!\left(\alpha+\frac{3}{2},\,\beta+1\right)
}{
B(\alpha+1,\,\beta+1)
}
\,{}_3F_2\!\left(
\frac12,\frac12,\alpha+\frac32;
\frac32,\alpha+\beta+\frac52;
1
\right).
\end{equation}
The physically relevant case of the above formulae for open quantum systems is $\alpha=n-2$ and $\beta=0$. Other physically relevant compact rank one symmetric spaces include the unit sphere $S^{n-1}$ in $\mathbb{R}^n$ (statistical mechanics and information theory), the real projective space $\mathbb{RP}^{n-1}$ (liquid crystals and topological defects), the quaternionic projective space $\mathbb{HP}^{n-1}$ (quantum field theory), and the Cayley plane $\mathbb{OP}^2$ (high-energy physics).

\section{Isotropic random walks on $\mathbb{CP}^{n-1}$}

\subsection{Fidelity distribution and ensemble relaxation}

First we compute the distribution and expectation of the fidelity defined as $F \equiv \cos^2(\vartheta) \in [0,1]$ for the complex projective space, where $\vartheta=d_{FS}(\xi_0,\xi)$ is the Fubini--Study distance of the initial and final state of the walker. Substituting $\alpha=n-2$ and $\beta=0$ into Eq. (\ref{pdf}) results in:
\begin{equation}
\label{tpdf}
\mathcal{F}_m(\vartheta) = 2\,\sin^{2n-3}(\vartheta)\,\cos(\vartheta)\,\sum_{k=0}^\infty c_k\,\hat{\mu}_k^m\,P_k^{n-2,0}(\cos(2\vartheta)) \,,
\end{equation}
where $P_k^{\alpha,\beta}(z)$ is the unnormalised Jacobi polynomial, and $c_k = (2k+n-1) {k+n-2 \choose n-2}$ is the normalised Plancherel measure. The probability distribution function of $F$ reads as:
\begin{equation}
\label{Fpdf}
\mathcal{G}_m(\varphi) = \mathcal{F}_m(\vartheta(\varphi))\,\left| \frac{d\vartheta(\varphi)}{d\varphi}\right| = (1-\varphi)^{n-2}\sum_{k=0}^\infty c_k \, \hat{\mu}_k^m \,P_k^{n-2,0}(2\,\varphi-1) \,,
\end{equation}
where we used the inverse transform $\vartheta(\varphi) = \cos^{-1}\sqrt{\varphi}$. $\mathcal{G}_m(\varphi)$ is already a radial function, and therefore it is normalised as $\int_0^1 \mathcal{G}_m(\varphi)\,d\varphi=1$. The expectation of $F$ reads as:
\begin{equation*}
\mathbb{E}_m(F) = \int_0^1 \varphi\,\mathcal{G}_m(\varphi)\,d\varphi = \int_0^{\pi/2} \cos^2(\vartheta)\,\mathcal{F}_m(\vartheta)\,d\vartheta = 2 \sum_{k=0}^\infty c_k\, \hat{\mu}_k^m \, H_k \,, 
\end{equation*}
where
\begin{equation*}
H_k = \int_{0}^{\pi/2} \sin^{2n-3}(\vartheta)\,\cos^3(\vartheta)\,P_k^{n-2,0}(\cos(2\vartheta))\,d\vartheta \,.
\end{equation*}
Using the orthogonality properties of Jacobi polynomials, $H_0 = (2\,n\,(n-1))^{-1}$ and $H_1=(2\,n\,(n+1))^{-1}$, while $H_{k \geq 2} = 0$, which yields
\begin{equation}
\label{expF}
\mathbb{E}_m(F) = \frac{1}{n} + \left(1-\frac{1}{n}\right)\hat{\mu}_1^m \,.
\end{equation}
Since the expected fidelity depends only on the lowest nontrivial Jacobi mode, whereas geometric observables associated with the Fubini--Study distance receive contributions from the full spectrum via the transition kernels $\hat{\mu}_k$, distinct step-size distributions may exhibit identical fidelity relaxation while producing substantially different geometric spreading on the projective Hilbert space. This suggests the possibility of nontrivial optimization problems relating fidelity-based and geometric measures of stochastic state-space exploration. Furthermore, Eq. (\ref{Fpdf}) implies
\begin{equation}
\label{betaF} \lim_{m \to \infty} \mathcal{G}_m(\varphi) = (n-1)(1-\varphi)^{n-2} \,,
\end{equation}
i.e., the fidelity follows the beta-distribution in the $m \to \infty$ limit with
\begin{equation}
\label{limitF}
\lim_{m \to \infty} \mathbb{E}_m(F) = \frac{1}{n} \,,
\end{equation}
independently of the step-size distribution.
 
\subsection{Short-time localization and long-time equilibration}

Let $\Theta$ denote the random step-size (or stride) of the walk. As the Maclaurin series of $J_k^{n-2,0}(\cos(2\vartheta))$ converges uniformly on $\vartheta \in [0,\pi/2]$ for any fixed $n$ and $k$, the single-step transition kernel can be re-written in terms of the even moments of $\Theta$ as follows:
\begin{equation}
\label{transition}
\hat{\mu}_k = \sum_{j=0}^\infty \frac{1}{(2j)!}\,\left(\frac{d^{2j}}{d\vartheta^{2j}}J_k^{n-2,0}(\cos(2\vartheta))\right)_{\vartheta=0}\,\mathbb{E}(\Theta^{2j}) = 1 - \frac{k(k+n-1)}{n-1}\,\mathbb{E}(\Theta^2) + \dots \,,
\end{equation}
where $\mathbb{E}(\Theta^{2j}) = \int_0^{\pi/2} \vartheta^{2j}\,d\mu(\vartheta)$. Let $d\mu(\vartheta)$ be any step-size distribution on $[0,\pi/2]$ that satisfies Lindeberg's condition \cite{Lindeberg1922,Feller1935}, and let $\mathbb{E}(\Theta^2) \ll 1$. In this case, $\mathbb{E}(\Theta^{2j}) = o(\mathbb{E}(\Theta^2))$ for any $j \geq 2$, and therefore the leading order of $\hat{\mu}_k$ is provided by exactly the first two terms of the sum in Eq. (\ref{transition}). Furthermore, the $m$-step transition kernels can be approximated as: 
\begin{equation}
\label{diff}
\hat{\mu}_k^m \approx \left(1 + \lambda_k\,\kappa \right)^m \approx \exp(\lambda_k\,\tau) \,,
\end{equation}
where $\lambda_k = -4\,k\,(k+n-1)$ is the $k$\textsuperscript{th} eigenvalue of the radial part of the Laplace--Beltrami operator corresponding to the Fubini--Study metric, $\kappa = \mathbb{E}(\Theta^2)/[4(n-1)]$, and 
\begin{equation*}
\tau = m\,\kappa = \frac{m\,\mathbb{E}(\Theta^2)}{4(n-1)} \,.
\end{equation*}
It is important to note that we use the standard Algebraic Geometric / Lie Theory definition of the Fubini--Study Laplace--Beltrami operator $\Delta_{FS}$, i.e., the maximal holomorphic sectional curvature is $H=4$. Using Eq. (\ref{diff}) in Eq. (\ref{tpdf}) results in:
\begin{equation*}
\mathcal{F}_m(\vartheta)|_{\mathbb{E}(\Theta^2) \ll 1} \approx 2\,\sin^{2n-3}(\vartheta)\,\cos(\vartheta)\,\sum_{k=0}^\infty c_k\,e^{\lambda_k \tau}\,P_k^{n-2,0}(\cos(2\vartheta)) \,,
\end{equation*}
which coincides with the radial heat kernel on $\mathbb{CP}^{n-1}$ under the map $\tau=D\,t$, where $D$ is the diffusion constant of the associated diffusion process \cite{Benabdallah1973,Demni2014} (the corresponding diffusion equation reads as $\partial_t p = D\,\Delta_{FS}\,p$). In other words, the limit process of an isotropic random walk satisfying Lindeberg's condition is an isotropic Brownian motion under the correspondence relation \cite{EthierKurtz,StroockVaradhan,Hsu2002}
\begin{equation}
\label{map}
4\,(n-1)\,D\,t = m\,\mathbb{E}(\Theta^2)\,.
\end{equation}
For $\tau \ll 1$, the radial heat kernel of the Brownian motion on $\mathbb{CP}^{n-1}$ can be approximated by that on $\mathbb{R}^{2(n-1)}$ with $\vartheta$ formally taking the role of the Euclidean distance \cite{Varadhan1967}. Accordingly, $\mathbb{E}(\vartheta^2)|_{\tau \ll 1} \approx 4\,(n-1)\,D\,t = m\,\mathbb{E}(\Theta^2)$, and
\begin{equation}
\label{quantum}
\mathbb{E}(\vartheta)|_{\tau \ll 1} \approx \frac{\Gamma\left(n-\frac{1}{2}\right)}{\Gamma(n-1)}\,\sqrt{4\,D\,t} = \frac{\Gamma\left( n - \frac{1}{2}\right)}{\Gamma(n-1)}\,\sqrt{\frac{m\,\mathbb{E}(\Theta^2)}{n-1}} \,, 
\end{equation}
where the pre-factor follows from the approximation with effective Euclidean dimension $q=2(n-1)$, i.e.\ $\Gamma((q+1)/2)/\Gamma(q/2) = \Gamma(n-1/2)/\Gamma(n-1)$.

The other limit we consider is when $\tau \to \infty$, which indicates $m \to \infty$ for finite $n$ and $\mathbb{E}(\Theta^2)$ under Eq. (\ref{map}). In this limit, the expectation of $\vartheta$ can be computed by substituting $\alpha=n-2$ and $\beta=0$ into Eq. (\ref{limtheta}), thus yielding:
\begin{equation}
\label{classical} 
\lim_{\tau \to \infty} \mathbb{E}(\vartheta) = \frac{\pi}{2} \left( 1- \frac{\Gamma\left(n-\frac{1}{2}\right)}{\sqrt{\pi}\,\Gamma(n)} \right) \,.
\end{equation}
Eq. (\ref{quantum}) describes the regime in which the walk remains localized in a small neighbourhood of the initial state in Fubini--Study geometry. In contrast, Eq. (\ref{classical}) corresponds to the asymptotic equilibration regime, where the walk approaches the uniform Haar measure and effectively loses memory of the initial state. The crossover regime can be estimated by equating $\mathbb{E}(\vartheta)|_{\tau \ll 1}$ and $\lim_{\tau \to \infty} \mathbb{E}(\vartheta)$, which provides a geometric criterion for the transition between localized and Haar-uniform evolution. Accordingly, let
\begin{equation*}
s^*(n) \equiv \frac{\pi}{2}\,\frac{1}{\sqrt{n-1}}\left( \frac{\Gamma(n)}{\Gamma(n-1/2)}- \frac{1}{\sqrt{\pi}}\right)
\end{equation*}
be the crossover parameter. For $\sqrt{m\,\mathbb{E}(\Theta^2)} \ll s^*(n)$, the walk remains localized in a small neighbourhood of the initial state in Fubini--Study geometry, corresponding to a coherence-preserving regime. In contrast, $\sqrt{m\,\mathbb{E}(\Theta^2)} \gg s^*(n)$ indicates an asymptotic equilibration regime in which recoverable phase coherence relative to the initial state is effectively lost.

\subsection{Mean first passage time and first passage time distribution}

From the viewpoint of quantum information processing, an important question is how long it takes for a stochastic evolution to reach a prescribed neighbourhood of a target state starting from a fixed initial state. In the theory of stochastic processes, this is known as the first passage time problem. Accordingly, let $\eta \in \mathbb{CP}^{n-1}$ be a target point in the complex projective space. In the context of random walks, the first passage time is defined as the smallest number of steps required for the walker to reach the target set $B_\varepsilon(\eta) \equiv \{ \xi \in \mathbb{CP}^{n-1}: d(\xi,\eta) < \varepsilon \}$:
\begin{equation*}
M \equiv \inf(m \geq 0 : d(\xi_m, \eta) \leq \varepsilon).
\end{equation*}
As the problem is radially symmetric, the mean first passage time $m^\ast(\vartheta) \equiv \mathbb{E}(M)$ is only a function of the distance of the initial and the target state, $\vartheta \equiv d(\xi_0,\eta)$, and satisfies the backward renewal equation
\begin{equation}
\label{MFPT}
m^\ast = 1 + P[m^\ast]
\end{equation}
on $\vartheta \in [0,\pi/2]$ with absorbing condition $m^\ast(\vartheta)=0$ on $\vartheta \in [0,\varepsilon]$. The transition probability operator reads as:
\begin{equation}
\label{Bwdop}
P[f](\vartheta)=\int_0^{\pi/2} \mathcal{K}_\mu(\vartheta,\lambda)f(\lambda)\,d\lambda \,,
\end{equation}
where the transition kernel is radial:
\begin{equation*}
\mathcal{K}_\mu(\vartheta, \lambda) = 2 \sin^{2n-3}(\lambda) \cos(\lambda)
\sum_{k=0}^{\infty}
 c_k\,\hat\mu_k\,J_k^{n-2,0}(\cos(2\vartheta))\,J_k^{n-2,0}(\cos(2\lambda)).
\end{equation*}
Although Eq.(\ref{MFPT}) is a Fredholm integral equation of the second kind, and the kernel is separable, the infinite number of Jacobi modes and the non-orthogonality of the eigenfunctions of $P$ for $\epsilon>0$ emerging from the killing condition $m^*(\vartheta)=0$ on $\vartheta \in [0,\epsilon]$ prevent an exact closed-form solution. In the Brownian limit, however, where $\hat\mu_k \approx 1 + \lambda_k\,\kappa$, the operator $\kappa^{-1}(P-I)$ converges to the radial part of $\Delta_{FS}$, which reads as:
\begin{equation}
\label{FSradial}
\mathcal{L} \equiv \frac{d^2}{d\vartheta^2} + \left[(2n-3)\cot(\vartheta)-\tan(\vartheta)\right]\frac{d}{d\vartheta}\,.
\end{equation}
Eq. (\ref{MFPT}) then reduces to the boundary-value problem \cite{Benabdallah1973,Demni2014,Grigoryan2009}
\begin{equation}
\label{int2ode}
\kappa\,\mathcal{L}[m^\ast] = -1
\end{equation}
on $\vartheta \in [\varepsilon,\pi/2]$ with absorbing boundary condition $m^\ast(\varepsilon)=0$ and regularity condition
\begin{equation*}
\frac{dm^\ast(\vartheta)}{d\vartheta}\bigg|_{\vartheta=\pi/2}=0.
\end{equation*}
The Neumann condition at $\vartheta=\pi/2$ reflects the symmetry of $\mathbb{CP}^{n-1}$: by isotropy, the radial flux of $m^\ast$ must vanish at the antipodal boundary, since there is no preferred direction away from that point. Eq. (\ref{int2ode}) is exactly solvable by the integrating factor method; the derivation is given in the Appendix. It is convenient to express the result in terms of fidelity \cite{Grigoryan2009}:
\begin{equation}
\label{MFPTsol}
m^\ast(F)=\frac{1}{\mathbb{E}(\Theta^2)}\left(
\ln\left|\frac{1-F}{1-F^\ast}\right|
-\sum_{k=1}^{n-2}
\frac{(1-F)^{-k}-(1-F^\ast)^{-k}}{k}
\right),
\end{equation}
where $F=\cos^2(d(\xi_0,\eta))$ is the fidelity between the initial and target state, and $F^\ast=\cos^2\varepsilon$ is the target fidelity. For $0 < 1-F^* \ll 1$, one obtains the logarithmic divergence $m^\ast(F)\propto -\ln(1-F^\ast)$ for $n=2$, while for $n>2$,
\begin{equation}
\label{striking}
m^\ast(F)\propto (1-F^\ast)^{2-n},
\end{equation}
indicating a much stronger divergence for high-fidelity targets. This behaviour is a direct consequence of the concentration properties of the Fubini--Study geometry: high-fidelity neighbourhoods occupy an exponentially vanishing fraction of the total volume of $\mathbb{CP}^{n-1}$, so the mean time to reach them grows without bound as $F^\ast \to 1$.

As isotropic Brownian motion on $\mathbb{CP}^{n-1}$ is exactly solvable, the distribution of the first passage time of the random walk can also be approximated analytically. Let $\zeta_t$ be an isotropic  Brownian motion on $\mathbb{CP}^{n-1}$ with diffusion constant $D$ and let
\begin{equation*}
T \equiv \inf(t \geq 0 : d(\zeta_t,\eta) \leq \varepsilon)
\end{equation*}
be the first passage time of $\zeta_t$. For $d(\zeta_0,\eta) \leq \epsilon$, $T=0$. Let
\begin{equation*}
S_\varepsilon(t,F) \equiv \mathbb{P}(T>t\mid \cos^2(d(\zeta_0,\eta)) = F)
\end{equation*}
be the survival probability as a function of the initial fidelity $F \in [0,F^*]$. $S_\varepsilon(t,F)$ satisfies the diffusion equation
\begin{equation}
\label{survive}
\partial_t S_\varepsilon = D\,\mathcal{L}_F[S_\varepsilon]
\end{equation}
on $F \in [0,F^*]$ with initial condition $S_\varepsilon(0,F)=1$ on $F \in [0,F^*)$ and boundary conditions $(\partial S_\varepsilon(t,F)/\partial F)_{F=0}=0$ and $S_\varepsilon(t,F^\ast)=0$ for $t \geq 0$. Furthermore,
\begin{equation}
\label{Loperator}
\mathcal{L}_F = 4\,F\,(1-F)\,\frac{\partial^2}{\partial F^2} + 4\,(1-n\,F)\,\frac{\partial}{\partial F}
\end{equation}
is the radial Fubini--Study Laplacian expressed in terms of fidelity, which directly follows from Eq. (\ref{FSradial}). Eq. (\ref{survive}) can be solved by using separation of variables (the details of the derivation are provided in the Appendix). Accordingly, the first passage time probability density to initial fidelity $F$ and target fidelity $F^*$ reads as:
\begin{equation}
\label{FPpdf}
p(t|F,F^\ast) = -\frac{\partial S_\varepsilon(t,F)}{\partial t} = 4\,D\,\sum_{k=1}^{\infty} B_k\,\rho_k\,(\rho_k+n-1)\,{}_2F_1(-\rho_k,\rho_k+n-1,1,F)\,e^{-4\rho_k(\rho_k+n-1)\,D\, t}, \enskip
\end{equation}
where $\rho_k$ is the $k$\textsuperscript{th} positive real root of the equation
\begin{equation}
\label{rhoeq}
{}_2F_1(-\rho,\rho+n-1,1,F^\ast)=0,
\end{equation}
while the weights of the different modes emerge from the initial condition and read as
\begin{equation}
\label{Bcoeff}
B_k=
\frac{
\int_0^{F^\ast}
{}_2F_1(-\rho_k,\rho_k+n-1,1,F)(1-F)^{n-2}dF
}
{
\int_0^{F^\ast}
{}[_2F_1(-\rho_k,\rho_k+n-1,1,F)]^2(1-F)^{n-2}dF
} \,.
\end{equation}
It is important to note that $\rho=0$ is not a solution to Eq. (\ref{rhoeq}) for $F^*<1$, and therefore the sum in Eq. (\ref{FPpdf})  only runs for strictly positive roots: $0 < \rho_1 < \rho_2 < \dots$. This simply emerges from the fact the constant function is not an eigenfunction of $\mathcal{L}_F$ on $F \in [0,F^*]$ as it does not satisfy the boundary condition $S_\varepsilon(t,F^*)=0$.

As the random walk converges weakly to the Brownian motion if Lindeberg's condition applies, Eq. (\ref{FPpdf}) can be directly applied to approximate the first passage time distribution of the random walk through the correspondence relation described by Eq. (\ref{map}). For the choice of $D \equiv \mathbb{E}(\Theta^2)/[4(n-1)]$, $t$ is simply replaced by $m$ in Eq. (\ref{FPpdf}), thus yielding:
\begin{equation}
\label{FPpdfRW}
p(m|F,F^\ast) = \kappa_0\,\sum_{k=1}^{\infty} B_k\,\rho_k\,(\rho_k+n-1)\,
{}_2F_1(-\rho_k,\rho_k+n-1,1,F)\,e^{-\rho_k(\rho_k+n-1)\,\kappa_0 m} \,,
\end{equation}
where $\kappa_0=4\,D=\mathbb{E}(\Theta^2)/(n-1)$. For large $m$, $p(m|F,F^*)$ has an exponential tail:
\begin{equation}
\label{asymp}
p(m|F,F^\ast) \sim C(F,F^\ast)\exp[\lambda_1(F^\ast,n)\,\kappa\,m],
\end{equation}
where $\lambda_1(F^\ast,n)=-4\rho_1(\rho_1+n-1)$ is the principal eigenvalue of the killed Fubini-Study Laplace-Beltrami operator. As 
\begin{equation}
\label{principal}
|\lambda_1(F^*,n)| \sim F^*(n-1)\,(n-2)\,(1-F^*)^{n-2} \,,
\end{equation} 
the exponential tail of $p(m|F,F^*)$ becomes longer with increasing $n$, which accords with Eq. (\ref{striking}). For the details of the derivation of Eq. (\ref{principal}), see the Appendix.

\section{Comparison with quantum mechanics}

\subsection{Isotropic quantum depolarizing channel}

Equations (\ref{expF}) and (\ref{betaF}) were originally derived in the framework of quantum information theory, see e.g. \cite{ZyczkowskiSommers2005,BengtssonZyczkowski2006}. In standard quantum information theory, an isotropic $U(n)$-covariant noise model acting on an initial pure state $|\psi_0\rangle$ is represented by the depolarizing density matrix
\begin{equation}
\label{isochannel}
\rho
=
q|\psi_0\rangle\langle\psi_0|
+
(1-q)\frac{I}{n} \,,
\end{equation}
where $0<q<1$. For this model, the observable fidelity with respect to the initial state is
\begin{equation}
\label{depol}
\bar F
=
\langle\psi_0|\rho|\psi_0\rangle
=
q+(1-q)\frac1n.
\end{equation}
Comparing Equations (\ref{depol}) and (\ref{expF}) shows that the isotropic random walk reproduces the fidelity relaxation law under the identification $q=\hat{\mu}_1^m$. Although this might seem as merely a formal equivalence, the correspondence can be made explicit at the level of the reduced density matrix. Namely, let $x_m\in\mathbb{C}^n$, $\|x_m\|=1$, be a representative of $\xi_m$ in the embedding space, and define the ensemble-averaged density matrix by $\rho_m\equiv\mathbb{E}(x_m x_m^\dagger)$. Since $\xi_m$ is an isotropic walk, $\rho_m$ can be computed explicitly, and reads as (the details of the derivation can be found in the Appendix):
\begin{equation}
\label{isomatrix}
\rho_m
=
\hat{\mu}_1^{\,m}\,x_0x_0^\dagger
+
\left(1-\hat{\mu}_1^{\,m}\right)\frac{I}{n}  \,.
\end{equation}
Comparing Equations (\ref{isochannel}) and (\ref{isomatrix}) confirms that the proposed model reproduces the evolution of an isotropic quantum depolarizing channel under the identification $q=\widehat{\mu}_1^{\,m}$. In particular, the system becomes maximally mixed for $m \to \infty$: $\lim_{m \to \infty} \rho_m = I/n$. As isotropic quantum depolarizing channels are physically realisable, the isotropic projective random walk belongs to a class of stochastic dynamics whose ensemble behaviour is consistent with established physical quantum channels.

Before proceeding, it is important to emphasise that our results concern equilibration to the Haar measure and should not be directly compared with results on finite-temperature Gibbs thermalisation in physical systems. The relevance of the result lies in the methodology: While Equations (\ref{expF}) and (\ref{betaF}) are typically obtained using Haar integration, random-state methods, and unitary invariance arguments, in the present framework they arise naturally from the geometry of the complex projective space. From this perspective, the derivation is essentially immediate and extends naturally to finite-step transition kernels and dynamical quantities.

\subsection{Haar--typicality}

Eq. (\ref{Haareq}) implies that due to the decay of all nontrivial Jacobi modes, the isotropic walk on $\mathbb{CP}^{n-1}$ converges to the uniform Haar measure for $m \to \infty$. Accordingly, for an $N$-qubit system (for which $n=2^N$), Eq. (\ref{limitF}) yields
\begin{equation*}
\lim_{m \to \infty} \mathbb{E}_m(F)=2^{-N} \, ,
\end{equation*}
which reproduces the standard concentration-of-measure behaviour of Haar-random quantum states in high-dimensional Hilbert spaces \cite{ZyczkowskiSommers2005,HaydenLeungWinter2006}. Equations (\ref{quantum}) and (\ref{classical}) provide a more detailed insight into the short- and long-time behaviour of the model. From the perspective of quantum information processing, the short-time localization may be interpreted as a coherence-preserving regime, while the long-time equilibration corresponds to an effective decoherence regime in which recoverable phase coherence relative to the prepared state is operationally lost. The asymptotic regime of the isotropic walk also inherits the standard typicality properties of Haar-random quantum states. In particular, Page's theorem \cite{Page1993} implies that Haar-random pure states in a sufficiently small subsystem of a bipartite system become nearly maximally mixed (the corresponding reduced density matrix reads as $\rho \approx I/n$), while canonical-typicality results \cite{PopescuShortWinter2006} establish that typical pure states within suitably constrained Hilbert subspaces can exhibit thermal properties at the subsystem level.

\subsection{Mean first passage time and first passage time distribution }

Eq. (\ref{striking}) is a central result of the paper. The concentration properties of the Fubini--Study geometry imply a rapid suppression of the probability of reaching high-fidelity target states in large quantum systems. The rapid growth of the mean first-passage time with Hilbert-space dimension is relevant to several problems in quantum information theory, including measurement-based quantum state preparation and feedback control \cite{Fu2014}, measurement-induced entanglement generation \cite{Williams2008}, and the preparation of highly entangled many-body states \cite{Dorstel2026}. Although these processes generally involve structured dynamics rather than isotropic diffusion, our results provide a geometric reference model for assessing the efficiency of stochastic exploration in reaching high-fidelity target states and quantifying the potential benefits of measurement, feedback, and other directed quantum dynamics. Furthermore, recent trapped-ion experiments have reported first-passage-time distributions with exponentially decaying long-time tails, thus providing a physical context in which stochastic diffusion models of quantum dynamics are relevant~\cite{Ryan2026}. Moreover, while exponential tails are generic for killed diffusions on compact domains  \cite{Grigoryan2009}, the dependence of $\lambda_1(F^\ast,n)$ in Eq. (\ref{asymp}) on the target fidelity and Hilbert-space dimension is highly non-trivial. 

\section{Concluding remarks}

The results presented here establish isotropic random walks on complex projective space as an analytically tractable reference model for stochastic evolution on the quantum-state manifold. As the algebra of radial functions on a compact rank-one symmetric space carries the Jacobi hypergroup structure, harmonic analysis yields explicit transition kernels, fidelity statistics, and geometric spreading laws. The model naturally reproduces the reduced density matrix of an isotropic quantum depolarizing channel, while the principal new results concern the first-passage problem. In the Brownian limit, we obtain an explicit spectral representation of the first-passage-time distribution, Eq. (\ref{FPpdfRW}), and derive the high-fidelity asymptotics of the principal decay rate, Eq.~(\ref{principal}). The latter reveals how the long-time first-passage tail depends on the Hilbert-space dimension.

A notable feature of the construction is that these results follow from geometric and probabilistic considerations alone, without reference to microscopic system–environment Hamiltonians, Lindblad generators, or stochastic Schrödinger equations. Although this limits the model's direct applicability to physical systems, it isolates those properties of stochastic quantum-state evolution that arise from projective geometry rather than specific dynamical mechanisms. In particular, the fidelity statistics and first-passage results reveal a geometric limitation on stochastic quantum-state preparation in large systems: both the expected distance from the initial state and the mean first-passage time to a high-fidelity target increase with dimensionality.

The origin of this behaviour is particularly transparent in the Brownian limit, where the fidelity process is governed by a Jacobi diffusion \cite{KarlinTaylor1981,EthierKurtz,RevuzYor1999,Hsu2002,BaudoinWang2016}:
\begin{equation*}
dF_t = 4\,D\,(1-n\,F_t)\,dt+\sqrt{8\,D\,F_t(1-F_t)}\,dW_t \,,
\end{equation*}
where $D$ is the diffusion coefficient and $W_t$ is a standard Wiener process on $\mathbb{R}$. The dimensional dependence of the long-time mean fidelity and first-passage statistics is encoded in the It\^o correction (geometric drift), reflecting the geometry of $\mathbb{CP}^{n-1}$. Thus, isotropic stochastic exploration becomes increasingly inefficient in large Hilbert spaces in the absence of additional dynamical mechanisms.

We emphasize that the present construction is intentionally minimal. Applications to realistic open quantum systems may require several extensions. Microscopic system–bath models show that time-dependent driving can produce non-Markovian responses and memory effects \cite{Gamba2025,GrabertThorwart2018}, while nonlinear coupling may lead to spatially inhomogeneous damping and diffusion \cite{MassignanLampoWehrLewenstein2015}. More generally, escape and first-passage kinetics can depend sensitively on environmental friction \cite{HanggiTalknerBorkovec1990,PollakGrabertHanggi1989}. In repeatedly measured quantum systems, first-detected-passage statistics may exhibit measurement backaction, quantum Zeno effects, dark states, and interference \cite{FriedmanKesslerBarkai2017,ThielBarkaiKessler2018}, while restart protocols can modify hitting-time statistics \cite{YinBarkai2023,YinWangBarkai2024}. Incorporating Hamiltonian circulation, anisotropic or state-dependent dissipation, environmental memory, and measurement back-action into the present framework therefore constitutes a natural direction for future work. Such extensions should preserve sufficient analytical tractability to identify the effects of individual dynamical mechanisms and should satisfy the structural constraints of open-system dynamics and quantum information theory.

\section*{Acknowledgements}

The author acknowledges the stimulating research environment and scientific interactions provided through the Department of Mathematical Sciences, the QUEST project and the Department of Physics at Loughborough University, where parts of this work were presented and discussed. The author declares no conflict of interest. No funding was received for this work.

\section*{Declaration of generative AI use}

Generative artificial intelligence tools (OpenAI ChatGPT and Anthropic Claude) were used as an aid in the preparation of this manuscript, including assistance with literature search, algebraic manipulations, and drafting and editing parts of the text. The AI-generated suggestions were critically assessed and independently verified by the author, and all final mathematical results and conclusions were determined and approved by the author. The author takes full responsibility for the content of the manuscript.

\section*{Data availability statement}

No new data were created or analysed in this study.

\bibliography{papers}

@article{Lindblad1976,
  author = {G. Lindblad},
  title = {On the generators of quantum dynamical semigroups},
  journal = {Communications in Mathematical Physics},
  volume = {48},
  pages = {119--130},
  year = {1976}
}

@article{BrodyHughston2001,
  author  = {Dorje C. Brody and Lane P. Hughston},
  title   = {Geometric quantum mechanics},
  journal = {Journal of Geometry and Physics},
  volume  = {38},
  number  = {1},
  pages   = {19--53},
  year    = {2001},
  doi     = {10.1016/S0393-0440(00)00052-8},
  issn    = {0393-0440},
  publisher = {Elsevier}
}

@article{Gorini1976,
  author = {V. Gorini and A. Kossakowski and E. C. G. Sudarshan},
  title = {Completely positive dynamical semigroups of N-level systems},
  journal = {Journal of Mathematical Physics},
  volume = {17},
  pages = {821--825},
  year = {1976}
}

@article{Dalibard1992,
  author = {J. Dalibard and Y. Castin and K. M{\o}lmer},
  title = {Wave-function approach to dissipative processes in quantum optics},
  journal = {Physical Review Letters},
  volume = {68},
  pages = {580--583},
  year = {1992}
}

@article{Dum1992,
  author = {R. Dum and P. Zoller and H. Ritsch},
  title = {Monte Carlo simulation of the master equation in quantum optics},
  journal = {Physical Review A},
  volume = {45},
  pages = {4879--4887},
  year = {1992}
}

@article{GisinPercival1992,
  author = {N. Gisin and I. C. Percival},
  title = {The quantum-state diffusion model applied to open systems},
  journal = {Journal of Physics A},
  volume = {25},
  pages = {5677--5691},
  year = {1992}
}

@book{BreuerPetruccione2002,
  author = {H.-P. Breuer and F. Petruccione},
  title = {The Theory of Open Quantum Systems},
  publisher = {Oxford University Press},
  year = {2002}
}

@book{BarchielliGregoratti2009,
  author = {A. Barchielli and M. Gregoratti},
  title = {Quantum Trajectories and Measurements in Continuous Time},
  publisher = {Springer},
  year = {2009}
}

@article{Daley2014,
  author = {A. J. Daley},
  title = {Quantum trajectories and open many-body quantum systems},
  journal = {Advances in Physics},
  volume = {63},
  pages = {77--149},
  year = {2014}
}

@book{Helgason1984,
  author    = {Sigurdur Helgason},
  title     = {Groups and Geometric Analysis: Integral Geometry, Invariant Differential Operators, and Spherical Functions},
  publisher = {Academic Press},
  year      = {1984},
  isbn      = {978-0123385804}
}

@article{Roesler_Voit_1999,
title={Partial Characters and Signed Quotient Hypergroups},
volume={51},
DOI={10.4153/CJM-1999-006-6},
number={1},
journal={Canadian Journal of Mathematics},
author={R{\"o}sler, Margit and Voit, Michael},
year={1999},
pages={96–116}}

@article{Pap_Voit_1998,
title={Edgeworth expansion on n-spheres and Jacobi hypergroups},
volume={58},
DOI={10.1017/S0004972700032378},
number={3},
journal={Bulletin of the Australian Mathematical Society},
author={Pap, Gyula and Voit, Michael},
year={1998},
pages={393–401}
}

@article{ZyczkowskiSommers2005,
  title = {Average fidelity between random quantum states},
  author = {\ifmmode \dot{Z}\else \.{Z}\fi{}yczkowski, Karol and Sommers, Hans-J\"urgen},
  journal = {Phys. Rev. A},
  volume = {71},
  issue = {3},
  pages = {032313},
  numpages = {11},
  year = {2005},
  month = {Mar},
  publisher = {American Physical Society},
  doi = {10.1103/PhysRevA.71.032313},
  url = {https://link.aps.org/doi/10.1103/PhysRevA.71.032313}
}

@book{BengtssonZyczkowski2006,
  author    = {Ingemar Bengtsson and Karol {\.Z}yczkowski},
  title     = {Geometry of Quantum States: An Introduction to Quantum Entanglement},
  publisher = {Cambridge University Press},
  address   = {Cambridge},
  year      = {2006},
  isbn      = {9780521814515}
}

@article{Benabdallah1973,
  author  = {A. I. Benabdallah},
  title   = {Noyau de diffusion sur les espaces homog\`enes compacts},
  journal = {Bulletin de la Soci\'et\'e Math\'ematique de France},
  volume  = {101},
  year    = {1973},
  pages   = {265--283},
  language = {French}
}

@article{Varadhan1967,
  author  = {S. R. S. Varadhan},
  title   = {On the behavior of the fundamental solution of the heat equation with variable coefficients},
  journal = {Communications on Pure and Applied Mathematics},
  volume  = {20},
  number  = {2},
  year    = {1967},
  pages   = {431--455},
  doi     = {10.1002/cpa.3160200210}
}

@article{Demni2014,
  author    = {Nizar Demni},
  title     = {Distributions of truncations of the heat kernel on the complex projective space},
  journal   = {Annales math\'ematiques Blaise Pascal},
  volume    = {21},
  number    = {2},
  pages     = {1--20},
  year      = {2014},
  doi       = {10.5802/ambp.339},
  url       = {https://www.numdam.org/articles/10.5802/ambp.339/},
  mrnumber  = {3322612}
}

@article{Watabe2022,
  author  = {Shohei Watabe and Michael Zach Serikow and Shiro Kawabata and Alexandre Zagoskin},
  title   = {Efficient Criteria of Quantumness for a Large System of Qubits},
  journal = {Frontiers in Physics},
  volume  = {9},
  pages   = {773128},
  year    = {2022},
  doi     = {10.3389/fphy.2021.773128},
  url     = {https://doi.org/10.3389/fphy.2021.773128}
}

@article{HaydenLeungWinter2006,
  author  = {Patrick Hayden and Debbie Leung and Andreas Winter},
  title   = {Aspects of Generic Entanglement},
  journal = {Communications in Mathematical Physics},
  volume  = {265},
  number  = {1},
  pages   = {95--117},
  year    = {2006},
  doi     = {10.1007/s00220-006-1535-6}
}

@article{PopescuShortWinter2006,
  author  = {Sandu Popescu and Anthony J. Short and Andreas Winter},
  title   = {Entanglement and the foundations of statistical mechanics},
  journal = {Nature Physics},
  volume  = {2},
  pages   = {754--758},
  year    = {2006},
  doi     = {10.1038/nphys444}
}

@article{Page1993,
  author  = {Don N. Page},
  title   = {Average Entropy of a Subsystem},
  journal = {Physical Review Letters},
  volume  = {71},
  number  = {9},
  pages   = {1291--1294},
  year    = {1993},
  doi     = {10.1103/PhysRevLett.71.1291}
}

@article{Zurek2003,
  author  = {Wojciech H. Zurek},
  title   = {Decoherence, einselection, and the quantum origins of the classical},
  journal = {Reviews of Modern Physics},
  volume  = {75},
  number  = {3},
  pages   = {715--775},
  year    = {2003},
  doi     = {10.1103/RevModPhys.75.715}
}

@book{Wiseman2009,
  author    = {Howard M. Wiseman and Gerard J. Milburn},
  title     = {Quantum Measurement and Control},
  publisher = {Cambridge University Press},
  address   = {Cambridge},
  year      = {2009},
  doi       = {10.1017/CBO9780511813948}
}

@book{EthierKurtz,
  author    = {Stewart N. Ethier and Thomas G. Kurtz},
  title     = {Markov Processes: Characterization and Convergence},
  publisher = {John Wiley \& Sons},
  address   = {New York},
  year      = {1986},
  series    = {Wiley Series in Probability and Mathematical Statistics},
  isbn      = {9780471081869},
  doi       = {10.1002/9780470316658}
}

@book{StroockVaradhan,
  author    = {Daniel W. Stroock and S. R. S. Varadhan},
  title     = {Multidimensional Diffusion Processes},
  publisher = {Springer},
  address   = {Berlin},
  year      = {1979},
  series    = {Grundlehren der mathematischen Wissenschaften},
  volume    = {233},
  doi       = {10.1007/978-3-662-21900-8}
}

@article{Koornwinder1973,
  author  = {Tom H. Koornwinder},
  title   = {The Addition Formula for Jacobi Polynomials and Spherical Harmonics},
  journal = {SIAM Journal on Applied Mathematics},
  volume  = {25},
  number  = {2},
  pages   = {236--246},
  year    = {1973},
  doi     = {10.1137/0125026}
}

@book{Grigoryan2009,
  author    = {Alexander Grigoryan},
  title     = {Heat Kernel and Analysis on Manifolds},
  publisher = {American Mathematical Society},
  address   = {Providence, RI},
  year      = {2009},
  series    = {AMS/IP Studies in Advanced Mathematics},
  volume    = {47}
}

@article{Feller1935,
  author  = {William Feller},
  title   = {On the Central Limit Theorem and the Law of Large Numbers},
  journal = {Transactions of the American Mathematical Society},
  volume  = {38},
  number  = {1},
  pages   = {71--79},
  year    = {1935},
  doi     = {10.2307/1989704}
}

@article{Lindeberg1922,
  author  = {J. W. Lindeberg},
  title   = {Eine neue Herleitung des Exponentialgesetzes in der Wahrscheinlichkeitsrechnung},
  journal = {Mathematische Zeitschrift},
  volume  = {15},
  number  = {1},
  pages   = {211--225},
  year    = {1922},
  doi     = {10.1007/BF01494395}
}

@book{Percival1998QSD,
  author    = {Ian Percival},
  title     = {Quantum State Diffusion},
  publisher = {Cambridge University Press},
  address   = {Cambridge},
  year      = {1998},
  isbn      = {978-0521620079}
}

@article{JacobsSteck2006,
  author       = {Kurt Jacobs and Daniel A. Steck},
  title        = {A Straightforward Introduction to Continuous Quantum Measurement},
  journal      = {Contemporary Physics},
  volume       = {47},
  number       = {5},
  pages        = {279--303},
  year         = {2006},
  doi          = {10.1080/00107510601101934},
  eprint       = {quant-ph/0611067},
  archivePrefix= {arXiv}
}

@Article{BaudoinWang2016,
author={Baudoin, Fabrice and Wang, Jing},
title={Stochastic areas, winding numbers and Hopf fibrations},
journal={Probability Theory and Related Fields},
year={2017},
month={Dec},
day={01},
volume={169},
number={3},
pages={977-1005},
issn={1432-2064},
doi={10.1007/s00440-016-0745-x},
url={https://doi.org/10.1007/s00440-016-0745-x}
}

@book{Hsu2002,
  author    = {Elton P. Hsu},
  title     = {Stochastic Analysis on Manifolds},
  publisher = {American Mathematical Society},
  series    = {Graduate Studies in Mathematics},
  volume    = {38},
  year      = {2002},
  isbn      = {9780821829882}
}

@book{RevuzYor1999,
  author    = {Daniel Revuz and Marc Yor},
  title     = {Continuous Martingales and Brownian Motion},
  edition   = {3},
  publisher = {Springer},
  address   = {Berlin},
  year      = {1999},
  series    = {Grundlehren der mathematischen Wissenschaften},
  volume    = {293},
  isbn      = {9783540643258}
}

@article{Ryan2026,
  author  = {Joseph M. Ryan and Simon Gorbaty and Thomas J. Kessler and Mitchell G. Peaks and Stephen W. Teitsworth and Crystal Noel},
  title   = {Experimental Measurement of Quantum First-Passage-Time Distributions},
  journal = {Physical Review Research},
  volume  = {8},
  number  = {2},
  pages   = {L022025},
  year    = {2026},
  doi     = {10.1103/PhysRevResearch.8.L022025}
}

@book{KarlinTaylor1981,
  author    = {Samuel Karlin and Howard M. Taylor},
  title     = {A Second Course in Stochastic Processes},
  publisher = {Academic Press},
  address   = {New York},
  year      = {1981},
  isbn      = {9780123986504}
}

@article{Gamba2025,
  author  = {Gamba, Daniele and Cui, Bingyu and Zaccone, Alessio},
  title   = {Open quantum systems with particle and bath driven by time-dependent fields},
  journal = {Physical Review A},
  volume  = {112},
  pages   = {012207},
  year    = {2025},
  doi     = {10.1103/w3vk-wx62}
}

@article{YinBarkai2023,
  author  = {Yin, R. and Barkai, E.},
  title   = {Restart Expedites Quantum Walk Hitting Times},
  journal = {Physical Review Letters},
  volume  = {130},
  pages   = {050802},
  year    = {2023},
  doi     = {10.1103/PhysRevLett.130.050802}
}

@article{YinWangBarkai2024,
  author  = {Yin, Ruoyu and Wang, Qingyuan and Barkai, Eli},
  title   = {Instability in the quantum restart problem},
  journal = {Physical Review E},
  volume  = {109},
  pages   = {064150},
  year    = {2024},
  doi     = {10.1103/PhysRevE.109.064150}
}

@article{GrabertThorwart2018,
  author  = {Grabert, Hermann and Thorwart, Michael},
  title   = {Quantum mechanical response to a driven {Caldeira--Leggett} bath},
  journal = {Physical Review E},
  volume  = {98},
  pages   = {012122},
  year    = {2018},
  doi     = {10.1103/PhysRevE.98.012122}
}

@article{MassignanLampoWehrLewenstein2015,
  author  = {Massignan, Pietro and Lampo, Aniello and Wehr, Jan and Lewenstein, Maciej},
  title   = {Quantum Brownian motion with inhomogeneous damping and diffusion},
  journal = {Physical Review A},
  volume  = {91},
  pages   = {033627},
  year    = {2015},
  doi     = {10.1103/PhysRevA.91.033627}
}

@article{HanggiTalknerBorkovec1990,
  author  = {H{\"a}nggi, Peter and Talkner, Peter and Borkovec, Michal},
  title   = {Reaction-rate theory: fifty years after {Kramers}},
  journal = {Reviews of Modern Physics},
  volume  = {62},
  number  = {2},
  pages   = {251--341},
  year    = {1990},
  doi     = {10.1103/RevModPhys.62.251}
}

@article{PollakGrabertHanggi1989,
  author  = {Pollak, Eli and Grabert, Hermann and H{\"a}nggi, Peter},
  title   = {Theory of activated rate processes for arbitrary frequency dependent friction: Solution of the turnover problem},
  journal = {The Journal of Chemical Physics},
  volume  = {91},
  number  = {7},
  pages   = {4073--4087},
  year    = {1989},
  doi     = {10.1063/1.456837}
}

@article{ThielBarkaiKessler2018,
  author  = {Thiel, Felix and Barkai, Eli and Kessler, David A.},
  title   = {First Detected Arrival of a Quantum Walker on an Infinite Line},
  journal = {Physical Review Letters},
  volume  = {120},
  pages   = {040502},
  year    = {2018},
  doi     = {10.1103/PhysRevLett.120.040502}
}

@article{FriedmanKesslerBarkai2017,
  author  = {Friedman, H. and Kessler, D. A. and Barkai, E.},
  title   = {Quantum walks: The first detected passage time problem},
  journal = {Physical Review E},
  volume  = {95},
  number  = {3},
  pages   = {032141},
  year    = {2017},
  doi     = {10.1103/PhysRevE.95.032141}
}

@book{Helgason1999,
  author    = {Helgason, Sigurdur},
  title     = {The Radon Transform},
  edition   = {2},
  series    = {Progress in Mathematics},
  volume    = {5},
  publisher = {Birkh{\"a}user},
  address   = {Boston},
  year      = {1999},
  note      = {Chapter III, Section 2: Compact Two-point Homogeneous Spaces},
  doi       = {10.1007/978-1-4757-1463-0},
  isbn      = {978-0-8176-4109-2}
}

@article{Fu2014,
  author  = {Fu, Shuangshuang and
             Shi, Guodong and
             Proutiere, Alexandre and
             James, Matthew R.},
  title   = {Feedback policies for measurement-based
             quantum state manipulation},
  journal = {Physical Review A},
  volume  = {90},
  number  = {6},
  pages   = {062328},
  year    = {2014},
  doi     = {10.1103/PhysRevA.90.062328}
}

@article{Williams2008,
  author  = {Williams, Nathan S. and
             Jordan, Andrew N.},
  title   = {Entanglement genesis under continuous
             parity measurement},
  journal = {Physical Review A},
  volume  = {78},
  number  = {6},
  pages   = {062322},
  year    = {2008},
  doi     = {10.1103/PhysRevA.78.062322}
}

@article{Dorstel2026,
  author  = {D{\"o}rstel, T. and
             Iadecola, T. and
             Wilson, J. H. and
             Buchhold, M.},
  title   = {Frustration-Free Control and Absorbing-State
             Transport in Entangled State Preparation},
  journal = {Physical Review Letters},
  volume  = {136},
  number  = {12},
  pages   = {120603},
  year    = {2026},
  doi     = {10.1103/zp25-ppgk}
}

\appendix

\section*{Appendix}

\subsection*{Brownian limit of the backward renewal equation }

Using the orthogonality property of Jacobi polynomials, the $k$\textsuperscript{th} eigenfunction of the operator $P[.]$ defined by Eq. (\ref{Bwdop}) is the normalised radial Jacobi polynomial $J_k^{n-2,0}(\cos(2\,\vartheta))$, while the corresponding eigenvalue is $\hat{\mu}_k$. In the Brownian limit, $\hat{\mu}_k \approx 1 + \lambda_k\,\kappa$, and therefore
\begin{equation}
\label{App1}
P[J_k^{n-2,0}](\vartheta) = \hat{\mu}_k\,J_k^{n-2,0}(\vartheta) \approx (1+\lambda_k\,\kappa)\,J_k^{n-2,0}(\vartheta) \,.
\tag{A.1}
\end{equation}
Re-arranging Eq. (\ref{App1}) yields:
\begin{equation*}
\kappa^{-1}(P-I)[J_k^{n-2,0}](\vartheta)  \approx \lambda_k\,J_k^{n-2,0}(\vartheta) \,,
\end{equation*}
which implies that the operator $\kappa^{-1}(P-I)[.]$ converges to the radial part of the Fubini--Study Laplace-Beltrami operator.

\subsection*{Mean first passage time in the Brownian limit}

Eq.~(\ref{int2ode}) can be solved by using the integrating factor method. Let $\omega(\vartheta)\equiv dm^\ast(\vartheta)/d\vartheta$. The integrating factor of the first-order ODE for $\omega(\vartheta)$ reads as:
\begin{equation*}
\mu(\vartheta) = \exp\!\left(\int \left[(2n-3)\cot(\vartheta)-\tan(\vartheta)\right]d\vartheta\right) = \sin^{2n-3}(\vartheta)\,\cos(\vartheta) \,,
\end{equation*}
which results in
\begin{equation*}
\omega(\vartheta) = -\frac{1}{\kappa\,\mu(\vartheta)}\int_{\vartheta_0}^{\vartheta} \mu(\lambda)\,d\lambda = \frac{1}{\kappa\,\sin^{2n-3}(\vartheta)\,\cos(\vartheta)} \left[ C - \frac{\sin^{2(n-1)}(\vartheta)}{2(n-1)} \right] \,.
\end{equation*}
The regularity condition $\omega(\pi/2)=0$ yields $C=1/[2(n-1)]$. Integrating once more and using the absorbing boundary condition $m^\ast(\varepsilon)=0$ gives:
\begin{equation}
\label{formalm}
m^\ast(\vartheta) = \frac{1}{2\,\kappa\,(n-1)}\int_\varepsilon^\vartheta \frac{1-\sin^{2(n-1)}(\lambda)}{\sin^{2n-3}(\lambda)\,\cos(\lambda)}\,d\lambda \,.
\tag{A.2}
\end{equation}
The integral is evaluated by the substitution $y \equiv \sin^2\lambda$, $dy = 2\sin(\lambda)\cos(\lambda)\,d\lambda$, which yields
\begin{equation}
\label{formalm2}
\frac{1-\sin^{2(n-1)}(\lambda)}{\sin^{2n-3}(\lambda)\,\cos(\lambda)}\,d\lambda = \frac{1}{2}\,\frac{1-y^{n-1}}{y^{n-1}(y-1)}\,dy = \frac{1}{2} \sum_{k=1}^{n-1} y^{-k}\,dy \,.
\tag{A.3}
\end{equation}
Substituting Eq.~(\ref{formalm2}) into Eq.~(\ref{formalm}) yields:
\begin{equation}
\label{almost}
m^\ast(\vartheta) = \frac{1}{4\,\kappa\,(n-1)} \left[ \ln\left|\frac{\sin(\vartheta)}{\sin(\varepsilon)}\right|^2 - \sum_{k=2}^{n-1} \frac{\sin^{2(1-k)}(\vartheta)-\sin^{2(1-k)}(\varepsilon)}{k-1} \right] \,.
\tag{A.4}
\end{equation}
Finally, Eq.~(\ref{MFPTsol}) is obtained from Eq.~(\ref{almost}) by substituting $F=\cos^2(\vartheta)$, $F^\ast=\cos^2(\varepsilon)$, hence $1-F=\sin^2(\vartheta)$, $1-F^\ast=\sin^2(\varepsilon)$, together with $\kappa=\mathbb{E}(\Theta^2)/[4(n-1)]$.

\subsection*{First passage time distribution in the Brownian limit}

Substituting $S_{\varepsilon}(t,F) \equiv f(t)\,g(F)$ into Eq. (\ref{survive}) results in
\begin{equation}
\label{FPdistApp1}
\frac{1}{D}\,\frac{f'}{f} = \frac{\mathcal{L}_F[g]}{g} \,,
\tag{A.5}
\end{equation}
where $\mathcal{L}_F[.]$ is defined by Eq. (\ref{Loperator}). As the left-hand side of Eq. (\ref{FPdistApp1}) depends only on $t$, and the right-hand side on $F$, they must be constant. Let this constant be $\lambda$, where $\lambda \in \mathbb{R}$. This yields $f(t) = e^{D\,\lambda\,t}$, while the spatial eigenvalue problem becomes
\begin{equation}
\label{FPdistApp2}
\mathcal{L}_F[g] - \lambda\,g = 0 \,.
\tag{A.6}
\end{equation}
Introducing the new parameter $\rho$ via $\lambda = -4\, \rho\,(\rho+n-1)$, dividing by $4$ ,then comparing Eq. (\ref{FPdistApp2}) to the Gauss hypergeometric equation
\begin{equation}
\label{FPdistApp6}
F\,(1-F)\,g'' +[c-(a+b+1)\,F]\,g'-a\,b\,g = 0
\tag{A.7}
\end{equation}
yields the parameters $c=1$, $a=-\rho$, and $b=\rho+n-1$. The solution of Eq. (\ref{FPdistApp6}) regular at $F=0$ reads as $g(F)={}_2F_1 (a,b,c,F)={}_2 F_1 (-\rho,\rho+n-1,1,F)$, thus yielding 
\begin{equation*}
S_\varepsilon(t,F) = e^{-4\,\rho\,(\rho+n-1)\,D\,t}\,{}_2F_1(-\rho,\rho+n-1,1,F) \,.
\end{equation*}
The absorbing condition $S_\varepsilon(t,F^*)=0$ yields the condition
\begin{equation}
\label{FPdistApp3}
{}_2F_1(-\rho,\rho+n-1,1,F^*)=0 \,.
\tag{A.8}
\end{equation}
Eq. (\ref{FPdistApp3}) has infinitely many strictly positive real roots $0 < \rho_1 < \rho_2 < \dots$ for $F^* \in (0,1)$. Accordingly, the full solution of Eq. (\ref{survive}) reads as:
\begin{equation*}
S_\varepsilon(t,F) = \sum_{k=1}^\infty B_k\,e^{-4\,\rho_k\,(\rho_k+n-1)\,D\,t}\,{}_2 F_1 (-\rho_k,\rho_k+n-1,1,F) \,.
\end{equation*}
The coefficients $B_k$ can be computed from the initial condition as follows. Introducing $\phi_k(F) \equiv {}_2F_1(-\rho_k,\rho_k+n-1,1,F)$, the orthogonality property reads as
\begin{equation}
\label{FPdistApp4}
\int_0^{F^*} \phi_k(F)\,\phi_l(F)\,(1-F)^{n-2}\,dF = \delta_{kl} \,.
\tag{A.9}
\end{equation}
Furthermore, the initial condition reads as
\begin{equation}
\label{FPdistApp5} 
S_\varepsilon(0,F)=\sum_{k=1}^\infty B_k\,\phi_k(F)=1
\tag{A.10}
\end{equation}
on $F \in [0,F^*)$. Multiplying Eq. (\ref{FPdistApp5}) by $\phi_l(F)$, integrating from $0$ to $F^*$, and using Eq. (\ref{FPdistApp4}) directly yields Eq. (\ref{Bcoeff}).

\subsection*{Large-$n$ asymptotics of the principal eigenvalue}

For fixed $F^*\in(0,1)$, the principal root $\rho_1$ of Eq. (\ref{FPdistApp3}) converges to $0$ as $n\to\infty$. We therefore start with expanding the hypergeometric function about $\rho=0$:
\begin{equation*}
{}_2F_1(-\rho,\rho+n-1;1;F^*)
=
\sum_{j=0}^{\infty}
\frac{(-\rho)_j(\rho+n-1)_j}{(j!)^2}(F^*)^j
=
1-A_n(F^*)\,\rho+O(\rho^2),
\end{equation*}
where
\begin{equation*}
A_n(F^*)=\sum_{k=1}^{\infty} \frac{(n-1)_k}{k!}\,\frac{(F^*)^k}{k} \,.
\end{equation*}
Consequently, the smallest positive root satisfies
\begin{equation*}
\rho_1
\sim
\frac{1}{A_n(F^*)}.
\end{equation*}
Consider now the Maclaurin series $(1-F)^{-a}=\sum_{k=0}^\infty \frac{(a)_k}{k!}\,F^k$. Setting $a=n-1$, subtracting $1$, dividing through by $F$, then integrating from $F=0$ to $F=F^*$ results in the integral representation 
\begin{equation}
\label{FPdistApp9}
A_n(F^*) = \int_{0}^{F^*}\frac{(1-F)^{-(n-1)}-1}{1-F}\,dF =  \int_{1-F^*}^{1}\frac{y^{-(n-1)}-1}{y}\,dy\,.
\tag{A.11}
\end{equation}
The integrand of the second integral in Eq. (\ref{FPdistApp9}) is a finite geometric series: $\frac{y^{-(n-1)}-1}{y} = \sum_{k=1}^{n-1} y^{-k}$, which yields 
\begin{equation*}
A_n(F^*)
= \int_{1-F^*}^1 \sum_{k=1}^{n-1}y^{-k}\,dy = -\log (1-F^*)
+
\sum_{k=1}^{n-2}
\frac{(1-F^*)^{-k}-1}{k} \,.
\end{equation*}
For fixed $F^*\in(0,1)$ and large $n$, the exponentially increasing
terms $(1-F^*)^{-k}$ dominate the sum. Setting $m=n-2$ gives
\begin{equation*}
A_n(F^*) \approx \sum_{k=1}^{m}\frac{(1-F^*)^{-k}}{k}
=
\frac{(1-F^*)^{-m}}{m}
\sum_{k=0}^{m-1}
\frac{(1-F^*)^k}{1-k/m} \approx \frac{(1-F^*)^{-m}}{F^*\,m} \,,
\end{equation*}
where we used $\lim_{m \to \infty} \sum_{k=0}^{m-1}\frac{(1-F^*)^k}{1-k/m} = \frac{1}{F^*}$. Consequently, $\rho_1 \sim F^*(n-2)\,(1-F^*)^{n-2}$. As $\rho_1$ is exponentially small as $n \to \infty$, $\rho_1\,(\rho_1+n-1) \sim (n-1)\rho_1$, and hence
\begin{equation*}
\rho_1\,(\rho_1+n-1)
\sim
F^*(n-1)\,(n-2)\,(1-F^*)^{n-2} \,.
\end{equation*}

\subsection*{Depolarizing channel for the random walk}

Eq. (\ref{isomatrix}) can be derived as follows. Suppose that the fidelity of $x_m$ relative to the initial state $x_0$ is $F_m$: $|\langle x_0,x_m \rangle|^2=F_m$. As $x_m$ is only a representative of $\xi_m$, it is indefinite up to a global phase factor $e^{\imath\,\varphi_m}$. Consequently, there exists $\varphi_m$ s.t. $x_m$ may be decomposed as:
\begin{equation*}
x_m
=
\sqrt{F_m}\,x_0
+
\sqrt{1-F_m}\,u_m ,
\end{equation*}
where $u_m\perp x_0$ and $\|u_m\|=1$ \cite[Sec.~4.5]{BengtssonZyczkowski2006}. As $F_m$ only fixes distance, $u_m$ is uniformly distributed on the unit sphere in the $(n-1)$-dimensional complex subspace orthogonal to $x_0$ \cite{Helgason1999}. Consequently,
\begin{equation}
\label{AppE1}
\mathbb{E}\!\left(u_m u_m^\dagger\mid F_m\right)
=
\frac{I-P_0}{n-1},
\tag{A.12}
\end{equation}
where $P_0=x_0x_0^\dagger$. The mixed terms vanish upon ensemble averaging,
\begin{equation}
\label{AppE2}
\mathbb{E}\!\left(x_0u_m^\dagger
+
u_mx_0^\dagger
\,\middle|\,F_m
\right)=x_0\,\mathbb{E}(u_m^\dagger|F_m) + \mathbb{E}(u_m|F_m)\,x_0^\dagger = 0 \,.
\tag{A.13}
\end{equation}
Using Equations (\ref{AppE1}) and (\ref{AppE2}), it follows that
\begin{equation*}
\mathbb{E}\!\left(x_mx_m^\dagger\mid F_m\right)
=
F_m P_0
+
\frac{1-F_m}{n-1}(I-P_0).
\end{equation*}
Defining the ensemble-averaged density matrix as
\begin{equation*}
\rho_m\equiv\mathbb{E}(x_mx_m^\dagger),
\end{equation*}
and averaging over the radial distribution gives
\begin{equation}
\label{AppE3}
\rho_m
=
\mathbb{E}_m(F)\,P_0
+
\frac{1-\mathbb{E}_m(F)}{n-1}(I-P_0).
\tag{A.14}
\end{equation}
Using Eq.~(\ref{expF}) in Eq. (\ref{AppE3}), we obtain Eq. (\ref{isomatrix}).

\end{document}